\documentclass[a4paper,10pt]{article}
\pdfoutput=1
\usepackage[utf8]{inputenc}
\usepackage[numbers]{natbib}
\usepackage{graphicx}
\usepackage[labelformat=simple]{caption,subcaption}
\usepackage{amsmath}
\usepackage{etoolbox}
\usepackage[overload]{empheq}
\usepackage{textcomp}
\usepackage{gensymb}
\usepackage{lineno}
\usepackage{caption}
\usepackage[margin=1in]{geometry}
\apptocmd{\sloppy}{\hbadness 10000\relax}{}{}
\title{Associative properties of structural plasticity based on firing rate homeostasis in recurrent neuronal networks}
\author{J\'ulia V. Gallinaro\textsuperscript{1*}, Stefan Rotter\textsuperscript{1}}

\graphicspath{{figures_v6/}{}}

\begin{document}

\maketitle
\begin{flushleft}
1 Bernstein Center Freiburg \& Faculty of Biology, University of Freiburg, Freiburg im Breisgau, Germany
\\
\bigskip
* julia.gallinaro@bcf.uni-freiburg.de
\\
\bigskip
\end{flushleft}

\begin{abstract}
Correlation-based Hebbian plasticity is thought to shape neuronal connectivity during development and learning, whereas homeostatic plasticity would stabilize network activity.
Here we investigate another, new aspect of this dichotomy: Can Hebbian associative properties also emerge as a network effect from a plasticity rule based on homeostatic principles on the neuronal level?
To address this question, we simulated a recurrent network of leaky integrate-and-fire neurons, in which excitatory connections are subject to a structural plasticity rule based on firing rate homeostasis.
We show that a subgroup of neurons develop stronger within-group connectivity as a consequence of receiving stronger external stimulation.
In an experimentally well-documented scenario we show that feature specific connectivity, similar to what has been observed in rodent visual cortex, can emerge from such a plasticity rule.
The experience-dependent structural changes triggered by stimulation are long-lasting and decay only slowly when the neurons are exposed again to unspecific external inputs.
\end{abstract}

\section*{Introduction}

Network plasticity involves connectivity changes at different levels.
Changes in the strength of already existing synapses are known as functional plasticity, whereas structural changes of axonal or dendritic morphology, as well as the creation of new and deletion of already existing synapses, is known as structural plasticity. During certain stages of development, axons and dendrites have been shown to grow and degenerate depending on neuronal activation \cite{Cohan1986, vanHuizen1987, Mattson1989}.
Structural changes, however, are not limited to the extent and shape of neurites, but also include more subtle alterations in spines and boutons.
Spine remodeling on excitatory cells due to neuronal activity has been observed \textit{in vitro} using organotypic hippocampal cultures \cite{Harris1999a, Wiegert2013, Oh2015} and \textit{in vivo} in sensory and in motor cortex of rodents \cite{Zuo2005b, Zuo2005, Hofer2009, Xu2009, Yang2009}.
Not only changes on postsynaptic spines, but also changes in presynaptic structures have been reported in organotypic hippocampal slice cultures \cite{Nikonenko2003,Schuemann2013, Ninan2006}.

The exact rules governing activity-dependent structural changes, however, are still not understood.
Computational models have tried to shed light on this issue by simulations, showing for example that many observed features of cortical connectivity could be achieved through the interaction of multiple plasticity mechanisms \cite{Zheng2014,Miner2016}, and that homeostatic regulation of neuronal activity on multiple time scales is necessary in order to stabilize Hebbian changes \cite{Zenke2015, Zenke2017}.
Homeostatic plasticity, in this context, usually refers to a regulation of neuronal connectivity that result in stabilization of neuronal activity at a set point.
There is growing evidence for homeostatic regulation of cortical connectivity (for a review, see Turrigiano \cite{Turrigiano2012}). 
More recently, homeostatic regulation of cortical activity has been demonstrated \textit{in vivo} in rodent visual cortex \cite{Keck2013, Hengen2013, Barnes2015, Hengen2016}. 
Hebbian plasticity, on the other hand, is used to describe mechanisms that change connections between two neurons based on the correlation between their respective activities. 
Many of these aspects were discussed at a recent conference devoted to the interaction of Hebbian and homeostatic plasticity \cite{Fox2017}.

But is it the case that the associative principles defining Hebbian learning must rely on pre-post correlations that are available exclusively at the level of individual synapses? 
Or could associative learning in networks also emerge from plasticity rules that are based on homeostatic principles on the level of whole neurons? 
The latter has been first proposed by Dammasch \cite{Dammasch1989} and, to our knowledge, has not been followed up since then. 
He proposed that Hebbian learning could emerge as a network property by using an algorithm based on firing rate homeostasis of individual neurons, with no reference to correlation.
The idea that associative learning could also emerge from the principle of homeostasis brings an important new aspect into the discussion of integrating Hebbian and homeostatic plasticity.
Experimental data usually describe the effects of plasticity on connectivity, but there are still many details missing.
Therefore, it is currently not an easy task to differentiate between three scenarios how associative learning might arise: It could emerge from a correlation-based Hebbian learning rule, it might arise as network effect of a learning rule based on homeostasis, or it could occur as a combination of both.

In this paper, we will explore these questions by testing the idea proposed by Dammasch \cite{Dammasch1989}, using a reimplementation of his algorithm in a more modern modeling framework.
We use a structural plasticity rule based on firing rate homeostasis recently implemented in NEST \cite{Butz2013, Diaz-Pier2016} and a recurrent network of leaky integrate-and-fire (LIF) excitatory and inhibitory neurons.
We show that a strongly interconnected assembly of neurons emerges when these neurons are jointly stimulated with stronger external input.
We then test the associative properties in an experimentally well-documented scenario, employing a simple model for the maturation of circuits in the primary visual cortex of rodents (V1).

It has been shown that neurons in adult V1 have an increased probability to be synaptically linked to other neurons that have a similar preference for visual features \cite{Ko2011}. 
Later, it was demonstrated that this feature-specific bias in connectivity was not present at the time of eye opening, and it developed only after some weeks of visual experience \cite{Ko2013}, suggesting that plastic mechanisms would shape the maturation of V1 circuits through visual experience.
Surprisingly, a feature-specific bias in connectivity was also shown to develop after eye opening in dark-reared mice lacking any visual input \cite{Ko2014}.
The idea of activity dependent plastic mechanisms shaping the maturation of these networks, however, was not ruled out.
Spontaneous retinal activity is also present on dark-reared animals \cite{Demas2003}, and plasticity could refine V1 networks based on patterned activity received by pairs of neurons that share some of these inputs.
Moreover, the relationship between connectivity and neuronal response to natural movies was not as strong for dark-reared mice as for normally reared mice \cite{Ko2014}, suggesting that plastic mechanisms contribute to the maturation of V1 circuits in an essential way, and the full maturation of feature specific connectivity would depend on visual experience anyway.

Recently, Sadeh \textit{et al.} \cite{Sadeh2015} have shown that a bias for feature specific connectivity can emerge in balanced random networks of LIF neurons with a synaptic plasticity rule that combines Hebbian and homeostatic mechanisms. 
They used, however, a functional plasticity rule, which limits a direct comparison to experimental data on connectivity. 
Here, we show that feature-specific connectivity can also emerge in a network of LIF neurons with a structural plasticity rule  in which correlations are implicitly evaluated through the random combination of presynaptic and postsynaptic elements in the network, and which does not require that synapses keep track of the presynaptic activity.
Moreover, we observed long-lasting structural after-effects of stimulation. 
This property is compatible with the notion of a persistent memory, which is not in every moment reflected by activity.

\section*{Methods}

\subsection*{Network simulations}
We simulate a recurrent network of $N=12\,500$ current-based LIF neurons, of which $N_{E}=0.8 N$ are excitatory and $N_{I}=0.2N$ are inhibitory.
The sub-threshold dynamics of the membrane potential $V_{i}$ of neuron $i$ obeys the differential equation
\begin{equation}
 \tau_{m} \frac{dV_{i}}{dt}=-V_{i}+ \tau_{m} \sum_{j} J_{ij} s_{j}(t-d),
\end{equation}
where $\tau_{m}$ is the membrane time constant.
The synaptic weight $J_{ij}$ from a presynaptic neuron $j$ to a postsynaptic neuron $i$ is the peak amplitude of the postsynaptic potential and depends on the type of the presynaptic neuron.
Excitatory connections have a strength of $J_{E}=J=0.1\,\mathrm{mV}$. Inhibitory connections are stronger by a factor $g=8$ such that $J_{I}=-gJ=-0.8\,\mathrm{mV}$.
A spike train $s_{j}(t)=\sum_k \delta (t-t_{j}^{k})$ consists of all spikes produced by neuron $j$. Some of these neurons represent the external input. 
The lumped spike train of all external neurons to a given neuron in the network is modeled as a Poisson process of rate $\nu_{ext}$, and external input to different neurons is assumed to be independent.
All synapses have a constant transmission delay of $d=1.5\,\mathrm{ms}$.
When the membrane potential reaches the firing threshold $V_\mathrm{th}=20\,\mathrm{mV}$, the neuron emits a spike to all postsynaptic neurons and its membrane potential is reset to $V_{r}=10\,\mathrm{mV}$ and held there for a refractory period of $t_\mathrm{ref}=2\,\mathrm{ms}$.

The indegree is fixed at $0.1 N_{I}$ for inhibitory to inhibitory and inhibitory to excitatory connections, and at $0.1 N_{E}$ for excitatory to inhibitory synapses. After connections of these types are established they remain unchanged throughout the simulation.
In contrast, excitatory to excitatory (EE) connections are initially absent and emerge only from the structural plasticity rule.
All simulations were conducted using the NEST simulator \cite{Gewaltig2007,Bos2015}. 
Numerical values of all parameters of our model are again collected in Table~\ref{tab:simulation_parameters}.

\begin{table}
 \caption{Parameters of the simulation and neuron model}
 \label{tab:simulation_parameters}
 \begin{tabular}{l c c}
  \hline
  \textbf{Parameter} & \textbf{Symbol} & \textbf{Value} \\
  \hline
  Number of neurons & $N$ & $12\,500$ \\
  \hline
  Number of excitatory neurons & $N_E$ & $10\,000$ \\
  \hline
  Number of inhibitory neurons & $N_I$ & $2\,500$ \\
  \hline
  Incoming excitatory connections per inhibitory neuron & $C_{E}$ & $1\,000$ \\
  \hline
  Incoming inhibitory connections per neuron & $C_{I}$ & $250$ \\
  \hline
  Reference weight & $J$ & $0.1\,\mathrm{mV}$ \\
  \hline
  Ratio inhibition to excitation& $g$ & $8$ \\
  \hline
  Excitatory weight & $J_{E}$ & $0.1\,\mathrm{mV}$ \\
  \hline
  Inhibitory weight & $J_{I}$ & $-0.8\,\mathrm{mV}$ \\
  \hline
  External weight & $J_\mathrm{ext}$ & $0.1\,\mathrm{mV}$ \\
  \hline
  Rate of external input & $\nu_\mathrm{ext}$ & $15\,\mathrm{kHz}$ \\
  \hline
  Membrane time constant & $\tau_{m}$ & $20\,\mathrm{ms}$ \\
  \hline
  Synaptic delay & $d$ & $1.5\,\mathrm{ms}$ \\
  \hline
  Threshold potential & $V_\mathrm{th}$ & $20\,\mathrm{mV}$ \\
  \hline
  Reset potential & $V_{r}$ & $10\,\mathrm{mV}$ \\
  \hline
  Refractory period & $t_\mathrm{ref}$ & $2\,\mathrm{ms}$ \\
  
  \hline
 \end{tabular}
\end{table}

\subsection*{Homeostatic structural plasticity (SP)}
The SP model used in our work has been recently implemented in NEST \cite{Diaz-Pier2016}.
The implementation combines precursor models by Dammasch \cite{dammasch1986self}, van Ooyen \& van Pelt \cite{VanOoyen1994} and van Ooyen \cite{vanOoyen1995}.
This model has been employed before to study the rewiring of networks after lesion or stroke \cite{Butz2009,Butz2013,Butz2014}, the specific properties of small-world networks \cite{Butz2014b}, the emergence of critical dynamics in developing neuronal networks \cite{Tetzlaff2010}, and neurogenesis in the adult dentate gyrus \cite{Butz2006,Butz2008inverse}.
All these models, however, included a distance-dependent kernel for the formation of new synapses, which is not part of the NEST implementation \cite{Diaz-Pier2016} that was used in our present study.

\subsubsection*{Neuronal activity and synaptic elements}
The EE connections in the network are volatile and undergo continuous remodeling, controlled by the SP algorithm. 
In its first versions, the model had continuous representations of pre- and postsynaptic densities, which were used for accessing connectivity between two neurons \cite{Butz2006}.
Later, it was adapted to have a discrete number of axonal and dendritic elements, which are combined to form synapses between neurons \cite{Butz2009}.
In the original model, the electrical activity of a neuron is represented by its intracellular calcium concentration, which is a lowpass filtered version of its time-dependent firing rate.
In this paper, we use the lowpass filtered spike train of neuron $i$ as a measure for its instantaneous firing rate $r_{i}$
\begin{equation}
 \tau_{r} \frac{dr_{i}(t)}{dt} + r_{i} = s_{i}(t).
\end{equation}
The time constant of the lowpass filter was chosen as $\tau_{r}=10\,\mathrm{s}$ throughout all our simulations.

Excitatory neurons are assigned a target rate $\rho$ and a set of pre- and postsynaptic elements, which can be interpreted as axonal boutons and dendritic spines, respectively. The number of presynaptic elements, $z^\mathrm{pre}$, and postsynaptic elements, $z^\mathrm{post}$, evolves in dependence of the neuron's firing rate according to
\begin{equation}
 \frac{dz^{k}_{i}(t)}{dt} = -\frac{1}{\beta^{k}} (r_{i}(t)-\rho), \qquad k\in\{\mathrm{pre}, \mathrm{post}\},
\end{equation}
where $i$ is the index of the neuron, and $\beta^{k}$ is a growth parameter.
We use a target rate $\rho=8\,\mathrm{Hz}$ and growth parameter $\beta=2$ for both pre- and postsynaptic elements of all excitatory neurons, unless stated otherwise.

Previous work on SP has used different functions to describe how these elements change with the neuron's activity, such as linear \cite{Butz2006,Butz2008inverse,Butz2009,Tetzlaff2010}, gaussian \cite{Butz2013,Butz2014,Diaz-Pier2016}, or logistic \cite{Butz2014b}.
Since there is currently no direct experimental data showing how the number of these elements vary with firing rate, we chose a generic linear function to implement a simple phenomenological model of firing rate homeostasis.
We use the same growth rule and the same parameters for both types of elements of all neurons in our network.
In the original model \cite{Butz2013}, free elements that are not engaged in a synapse will decay with a certain rate.
In the model considered here, however, free elements do not decay with time.
We did run test simulations considering the decay, and found that our main results were not altered (see Supplementary material).

At regular intervals $\Delta t=100\,\mathrm{ms}$, the structural plasticity rules are applied to delete already existing and create new synaptic contacts.
Numerical values of the parameters are summarized in Table~\ref{tab:plasticity_parameters}.

\begin{table}
 \caption{Parameters of the plasticity rule. Numbers in bold are the default values that are used if no specification is given.}
 \label{tab:plasticity_parameters}
 \begin{tabular}{l c c}
  \hline
  \textbf{Parameter} & \textbf{Symbol} & \textbf{Value} \\
  \hline
  Firing rate time constant & $\tau_{r}$ & $10 \, \mathrm{s}$ \\
  \hline
  Synaptic elements growth parameter & $\beta$ & $0.2,0.625,\textbf{2},6.25,20$ \\
  \hline
  Target rate & $\rho$ & $5,6,7,\textbf{8},9\,\mathrm{Hz}$ \\
  \hline
  Structural plasticity interval & $\Delta t$ & $100\,\mathrm{ms}$ \\
  \hline
 \end{tabular}
\end{table}

\subsubsection*{Synapse deletion and creation}
At regular intervals $\Delta t$, when rewiring is scheduled, a neuron may have more or less pre- and post-synaptic elements than it has actual synapses.
In that case, synapses are either deleted or created, in order to match the number of elements to the number of active synaptic contacts.
At fixed intervals $\Delta t$, the number of postsynaptic (presynaptic) elements is compared to the number of existing incoming (outgoing) synapses of each neuron
\begin{equation}
\Delta z^\mathrm{pre}_{j} = z^\mathrm{pre}_{j}-\sum_{i} C_{ij}
\qquad\text{and}\qquad
\Delta z^\mathrm{post}_{i} = z^\mathrm{post}_{i}-\sum_{j} C_{ij}, 
\end{equation}
where $C$ is the matrix containing the number of synapses between presynaptic neurons $j$ and postsynaptic neurons $i$. 
If the neuron has more synaptic contacts than synaptic elements ($\Delta z^{k}<0$), synapses are deleted.
The $|\Delta z^{k}|$ synapses to be deleted are randomly chosen among the existing contacts a neuron has.
After a synapse has been deleted due to a loss of presynaptic (postsynaptic) elements, the corresponding postsynaptic (presynaptic) elements remain available for a new connection.
If the neuron has more elements than contacts ($\Delta z^{k}>0$), the neuron is considered to have $\Delta z^{k}$ free synaptic elements.
All free synaptic elements in the network are randomly combined into pairs of pre- and postsynaptic elements to form new synapses.
The number of synapses formed is limited by both the total number of pre- and the total number of postsynaptic elements in the full network.
Each newly created synapse has a fixed strength $J$.
Multiple synapses between the same pair of neurons are allowed, but auto-synapses (self-connection of a neuron onto itself) are not.
See Butz \& van Ooyen \cite{Butz2013} and Diaz-Pier \textit{et al.} \cite{Diaz-Pier2016} for more details on the implementation of the model.

\subsection*{Subgroup stimulation}
The networks were first grown without structured input, with all excitatory neurons receiving external Poisson input with the same rate $\nu_\mathrm{ext}$.
Stimulation was started after $750\,\mathrm{s}$, when enough EE connections were grown, and, apart from small fluctuations, all neurons fired at their target rate.
During the stimulation period, a subgroup comprising $10\%$ of the excitatory neurons received an increased external input ($1.1\nu_\mathrm{ext}$ for $150\,\mathrm{s}$).
After stimulation, the external input was set back to its original value ($\nu_\mathrm{ext}$) for all excitatory neurons.
Both the activity and the connectivity of the network were monitored for $5\,500\,\mathrm{s}$.

\subsection*{Visual stimulation protocol}
For the visual cortex simulations, we consider a network similar to the one described in Sect. Network Simulations.
The stimulation only starts after the networks have created enough EE connections such that the preset firing rate can be maintained. Visual stimulation is simulated by providing the excitatory neurons with Poisson input the rate of which depends on the orientation of the stimulus, see Sadeh \textit{et al.} \cite{Sadeh2015} for details of the protocol.
The baseline firing rate $\nu_\mathrm{ext}$ is the same as during the growth period, the modulation depends on the orientation of the visual stimulation $\theta$ and the preferred orientation (PO) $\theta_\mathrm{PO}$ of the input and a modulation gain parameter $\mu$
\begin{equation}
\nu_\mathrm{mod}(\theta,\theta_\mathrm{PO})=\nu_\mathrm{ext}\left[1+\mu \cos(2(\theta-\theta_\mathrm{PO}))\right] .
\end{equation}

Each neuron is assigned a parameter $\theta_\mathrm{PO}$, which is randomly drawn from a uniform distribution on $[0\degree,180\degree)$.
During the stimulation phase, a different $\theta$ is randomly drawn from a uniform distribution on the interval $[0\degree,180\degree)$ and presented to all excitatory neurons for a duration of $t_\mathrm{st}=1\,\mathrm{s}$.
We use a modulation $\mu=0.15$, and the stimulation protocol consists of presenting a total of $N_\mathrm{st}=5\,000$ different stimuli. 
After the stimulation, the external input to all excitatory neurons is once again set to its initial non-modulated value of $\nu_\mathrm{ext}$, and the network is simulated for another $t_\mathrm{post}=10\,000\,\mathrm{s}$.
Numerical values of all parameters regarding the stimulation protocol are collected in Table~\ref{tab:stimulation_parameters}.

\begin{table}
 \caption{Parameters of the stimulation protocol}
 \label{tab:stimulation_parameters}
 \begin{tabular}{l l l}
  \hline
  \textbf{Parameter} & \textbf{Symbol} & \textbf{Value} \\
  \hline
  Modulation of external input & $\mu$ & $0.15$ \\
  \hline
  Stimulus orientation & $\theta$ & $[0\degree,180\degree)$ \\
  \hline
  Input preferred orientation & $\theta_\mathrm{PO}$ & $[0\degree,180\degree)$ \\
  \hline
  Time per stimulus & $t_\mathrm{st}$ & $1\,\mathrm{s}$ \\
  \hline
  Number of stimuli & $N_\mathrm{st}$ & $5\,000$ \\
  \hline
  Time post stimulation & $t_\mathrm{post}$ & $10\,000\,\mathrm{s}$ \\
  \hline
  
 \end{tabular}
\end{table}

\subsection*{Spike train analysis}
The spike count correlation between a pair of neurons $i$ and $j$ was calculated as the Pearson correlation coefficient
\begin{equation}
 R_{ij} = \frac{c_{ij}}{\sqrt{c_{ii}c_{jj}}},
\end{equation}
where $c_{ij}$ is the covariance between spike counts extracted from spike trains $x_{i}$ and $x_{j}$ of two neurons, and $c_{ii}$ is the variance of spike counts extracted from $x_{i}$. Correlations were calculated from spike trains comprising $20\,\mathrm{s}$ of activity, using bins of size $10\,\mathrm{ms}$.

The irregularity of spike train of neuron $i$ was measured as the coefficient of variation of its inter-spike intervals
\begin{equation}
 \mathrm{CV}_{i} = \frac{\sigma_{i}}{\mu_{i}},
\end{equation}
where $\mu_{i}$ is the mean and $\sigma_{i}$ is the standard deviation of the inter-spike intervals extracted from the spike train of neuron $i$ (duration $20\,\mathrm{s}$).

\section*{Data availability}
The datasets generated during and analysed during the current study are available from the corresponding author on reasonable request.

\section*{Results}
We start by growing recurrent networks of excitatory and inhibitory LIF neurons. 
All synaptic connections are static, except for EE connections, which are initially absent and grow according to a structural plasticity rule that implements firing rate homeostasis (see Methods for more details).
Before stimulation, we characterized the networks formed under the influence of the structural plasticity rule for uniform (untuned) external stimulation.

\subsection*{Grown networks are random in absence of structured input}
\begin{figure}
\centering
\makebox[\textwidth]{\includegraphics[scale=1]{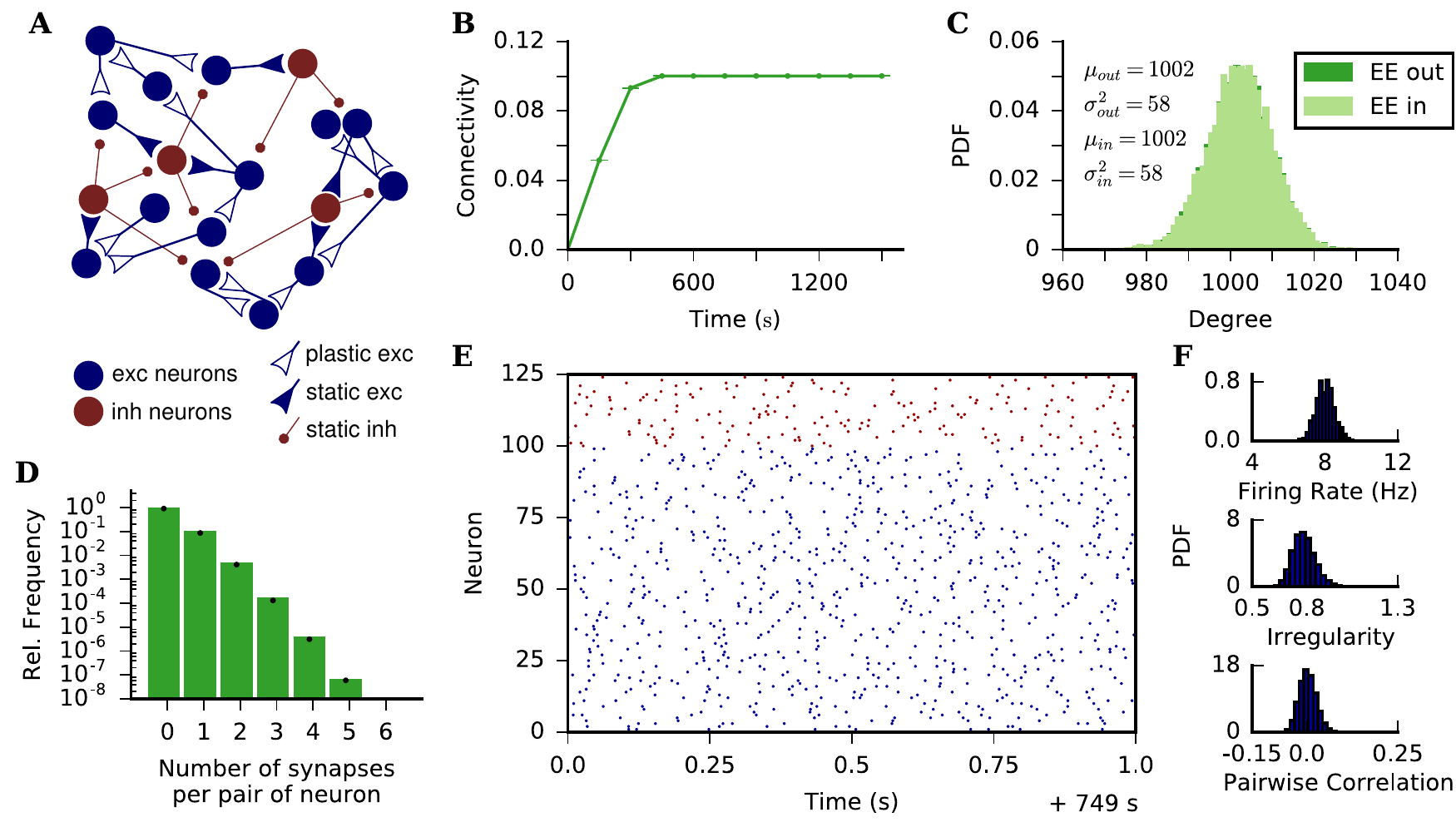}}
\caption{Structure of grown networks. (A)~The network is composed of $80\%$ excitatory and $20\%$ inhibitory LIF neurons. EE connections are plastic and follow the SP rule. All other connections are static and randomly created at the beginning of the simulation, such that all neurons have a fixed indegree corresponding to $10\%$ of the presynaptic population size. (B)~Time evolution of average EE connectivity. Dots and bars indicate mean $\pm$ standard deviation across $10$ independent simulation runs. Highest standard deviation in the time series is $2.8 \times 10^{-5}$. (C)~Indegree and outdegree distributions for EE connections after $750\,\mathrm{s}$ simulation time. $\mu_\mathrm{in}$, $\sigma^2_\mathrm{in}$, $\mu_\mathrm{out}$ and $\sigma^2_\mathrm{out}$ are the mean and the variance of the shown indegree and outdegree distributions, respectively.  (D)~Normalized histogram of the number of synapses per contact between pairs of neurons after $750\,\mathrm{s}$ of simulation. Black dots refer to a Poisson distribution with rate parameter matching the average connectivity of the simulated network. (E)~Raster plot showing $1\,\mathrm{s}$ activity of $1\,00$ excitatory and $25$ inhibitory neurons (randomly chosen) after $750\,\mathrm{s}$ of simulation. (F)~Histograms of firing rate, irregularity and pairwise correlation for all (pairs of) excitatory neurons after the network has reached a statistical equilibrium state (judged ``by eye''). The neurons fire in an asynchronous-irregular (AI) regime. Data were extracted from $20\,\mathrm{s}$ of activity, the bin size used for calculating $\mathrm{CC}$ was $10\,\mathrm{ms}$.  (B-F) Target rate $\rho=8\,\mathrm{Hz}$ and $\beta=2$ for all subplots.}
\label{fig:growth_structure} 
\end{figure}

The target rate is set to $\rho = 8\,\mathrm{Hz}$ for all excitatory neurons, which is the expected firing rate of excitatory neurons for the parameter set we are using and $10\%$ EE connectivity.
As expected, average EE in- and outdegree increase from $0$ until stabilizing at approximately $1\,000$, corresponding to an average EE connectivity of $10\%$ (Fig.~\ref{fig:growth_structure}B and C).
The plasticity rule is always active and individual EE connections are still being created and deleted, but we consider the network to be in equilibrium at this point, when excitatory neurons fire on average at their target rate.

Since multiple synapses between the same pair of neurons are allowed , we also looked into the distribution of the number of synapses between pairs of pre- and post-synaptic neurons.
Fig.~\ref{fig:growth_structure}D shows that this distribution is roughly a Poisson distribution.
If individual contacts can be considered as independent random variables with a Poisson distribution, the in- and outdegree of individual neurons would also follow a Poisson distribution and satisfy $\mu\approx\sigma^{2}$.
Fig.~\ref{fig:growth_structure}C shows, however, that the distribution of in- and outdegrees have $\sigma^{2} < \mu$.
In the SP model, in- and outdegree distributions change for different distribution of target rates.
We also performed simulations in which target rates were drawn from broader distributions, yielding also broader distributions of in- and outdegree,and the main results of feature specific connectivity were not altered (see Supplementary material).
Since a thorough study of the effect of target rate on degree distributions was beyond the scope of this paper, for simplicity we fixed $\rho=8\,\mathrm{Hz}$ for all excitatory neurons.

For the parameters used on our simulations, an inhibition dominated random network of LIF neurons with $10\%$ connection probability has been shown to have low firing rates, as well as asynchronous and irregular (AI) spike trains \cite{Brunel2000}.
Fig.~\ref{fig:growth_structure}E shows the activity of the network in equilibrium, after $750\,\mathrm{s}$ of simulation.
The population raster plot (Fig.~\ref{fig:growth_structure}E) indicates the network is in an AI state.
Pairs of neurons fire with low correlation coefficient ($\mathrm{CC}$, bin size $10\,\mathrm{ms}$) (Fig.~\ref{fig:growth_structure}F), and individual spike trains have a coefficient of variation ($\mathrm{CV}$) around $0.7$ (Fig.~\ref{fig:growth_structure}F).
As expected from the choice of $\rho$, the firing rate of excitatory neurons has a mean value of $8\,\mathrm{Hz}$ in equilibrium (Fig.~\ref{fig:growth_structure}F). 
Firing rates, pairwise correlation and irregularity were calculated from spike trains recorded during $20\,\mathrm{s}$ of simulation.

The networks grown according to the SP rule do exhibit some non-random features (see Supplementary Material). 
However, we consider these deviations from random networks as small. 
After $750\,\mathrm{s}$ of simulation, networks with a target rate $\rho=8\,\mathrm{Hz}$ for all excitatory neurons, and a growth parameter $\beta=2$ for all synaptic elements have an essentially random structure and an activity that can be classified as AI.

\subsection*{Time constant of growth process depends on target rate and the growth parameter for synaptic elements}
We then determined the time scale of network growth.
To that end we simulated a network in which all neurons receive Poisson input with the same rate $\nu_\mathrm{ext}$ until the equilibrium was reached.
We considered the network to be in equilibrium when connectivity (i.e.\ in- and outdegree distributions) are stable. 
In this state, individual synapses are still plastic and are created, deleted and recreated as the simulation runs.
For these simulations, all neurons had the same target firing rate $\rho$, but we simulated networks with different values set for $\rho$.

The evolution of EE connectivity is plotted in Fig.~\ref{fig:growth_structure}B.
The time scale of that growth process depends on how fast synaptic elements grow (parameter $\beta$), but also on the target rate $\rho$ ($\beta=2$ and $\rho=8\,\mathrm{Hz}$ on Fig.~\ref{fig:growth_structure}B).
The number of new synapses created in the network depends on the number of available free elements. 
This number, in turn, depends on the growth parameter $\beta$ for the synaptic elements.
Therefore, it comes as no surprise that the network growth also depends on $\beta$.
The observed relation between the time constant $\beta$ of network growth and target rate $\rho$, however, is not self-explaining, as for networks with fixed incoming inhibitory connections and fixed external input, the gain in firing rate depends non-linearly on the actual number of incoming excitatory connections to excitatory neurons \cite{Brunel2000}.

\begin{figure}
\centering
\makebox[\textwidth]{\includegraphics[scale=1]{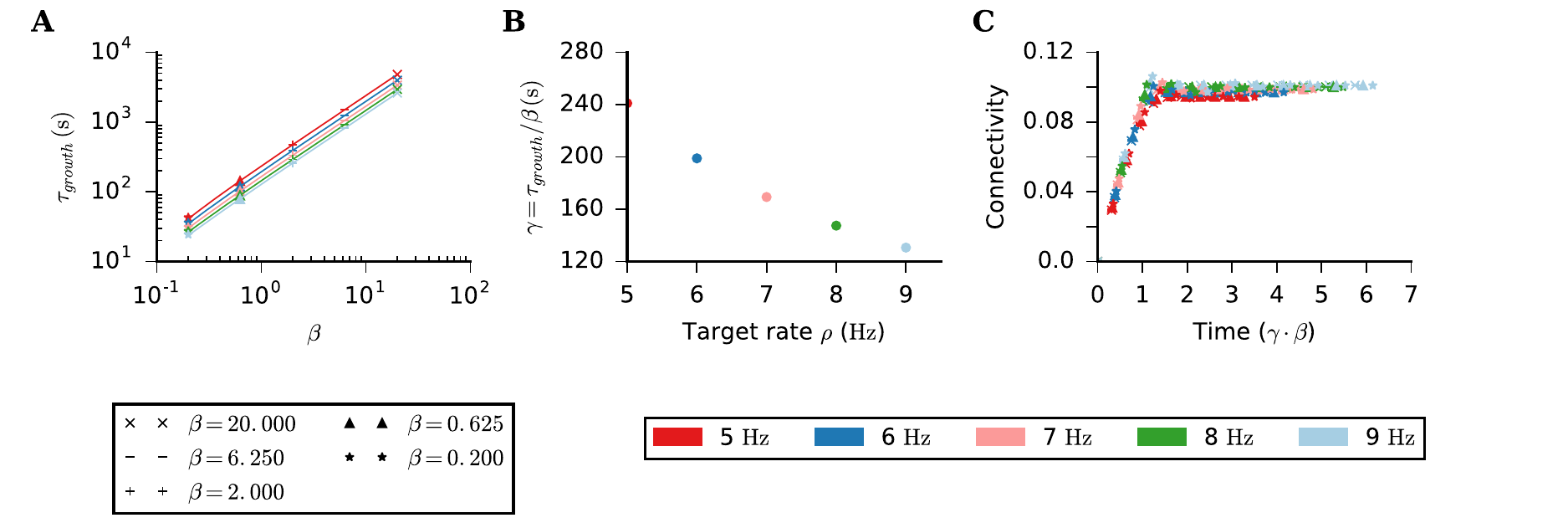}}
\caption{Time scale of network growth. (A)~Time constant of the growth process (defined in text) extracted from Fig.~\ref{fig:growth_structure}B plotted against $\beta$, for different values of $\rho$. This plot summarizes the result of $25$ simulation runs, using $5$ different values of $\beta$ and $5$ different values of $\rho$. In each simulation, all excitatory neurons have the same parameters $\rho$ and $\beta$, for both presynaptic and postsynaptic elements. (B)~Gain $\gamma=\tau_\mathrm{growth} / \beta$ plotted against $\rho$. (C)~Time evolution of the average EE connectivity of the $25$ simulation runs in (A), rescaled by $\tau_\mathrm{growth}$.} \label{fig:time_scale} 
\end{figure}

To better understand this dependence, we defined a growth time constant $\tau_\mathrm{growth}=s / \alpha$, where $s$ is the plateau value of the connectivity, calculated here as the average connectivity for the last $4$ discrete time points in a long enough simulation, and $\alpha$ is the highest (mostly initial) slope extracted from the connectivity time series.
We then plotted $\tau_\mathrm{growth}$ against the growth parameter $\beta$ of synaptic elements, for different values of $\rho$ (Fig.~\ref{fig:time_scale}A). The range of parameters considered here spanned more than $3$ orders of magnitude, so we used a log-log plot to represent it. The exponent, however, is $1$, indicating a linear dependency between $\beta$ and $\tau_\mathrm{growth}$.
Also, the growth of EE connectivity is faster for simulations with higher $\rho$, as can be seen in Fig.~\ref{fig:time_scale}A.
The time constant of EE connectivity growth is, therefore, a function of both $\beta$ and $\rho$.
In order to express this dependency of the growth time constant on $\rho$,we extracted the coefficient $\gamma=\tau_\mathrm{growth} / \beta$ from the simulated data (Fig.~\ref{fig:time_scale}B).
Finally we rescaled the time axis of Fig.~\ref{fig:growth_structure}B by dividing it by $\tau_\mathrm{growth}$ for given values of $\beta$ and $\rho$, obtaining a growth curve in time units of $\gamma \cdot \beta$ (Fig.~\ref{fig:time_scale}C).

We found that the growth process is stable throughout at least $3$ orders of magnitude of the parameter $\beta$.
An even slower process would lead to exceedingly long simulation times, but we have no reason to believe that this would destabilize the system.
A value of $\beta$ could easily be chosen such that network growth would happen in hours or days, matching experimental data of structural plasticity.
Making the process faster, however, may destabilize the system, as the rate of new contacts created would increase.
Apart from mathematical constraints, there are also biological limits regarding speed, of course, as a fast system also requires high turnover rates and efficient transport of proteins and other molecules.

From now on, all simulated networks have a target rate $\rho=8\,\mathrm{Hz}$ for all excitatory neurons, all synaptic elements have a growth parameter $\beta=2$. 

\subsection*{Associative properties of SP}
In his paper, Dammasch \cite{Dammasch1989} suggested that the compensation algorithm, on which this particular SP model \cite{Butz2013,Diaz-Pier2016} is based, could implement a form of Hebbian learning.
This becomes clear when analyzing the equations by Butz \& van Ooyen \cite{Butz2013} describing the expected change in connectivity induced by the SP algorithm:
\begin{equation}
\Delta C_{i,j}= 
\begin{cases}
C_{i,j}\frac{\Delta z^{pre}_{j}}{z^{pre}_{j}} & \text{for } \Delta z^{pre}_{j} < 0
\\
C_{i,j}\frac{\Delta z^{post}_{i}}{z^{post}_{i}} & \text{for } \Delta z^{post}_{i} < 0 
\\
\frac{\Delta z^{pre}_{j} \Delta z^{post}_{i}}{max(\sum_{k} \Delta z^{pre}_{k}, \sum_{k} \Delta z^{post}_{k})} & \text{for } \Delta z^{pre}_{j} > 0$ and $\Delta z^{post}_{i} > 0
\label{eq:delta_c_positive}
\end{cases} .
\end{equation}
Please note that we have adapted Eqs.~\ref{eq:delta_c_positive} from \cite{Butz2013} to match the nomenclature used in the present paper. Also, we only account for the simplest case here, in which only EE connections are plastic, and where the formation of synapses does not depend on distance.
Eq.~\ref{eq:delta_c_positive} clearly shows that the connectivity increase happens when two neurons are simultaneously in a low activity state, implementing Hebbian learning through a covariance rule.
On the other hand, in periods of elevated activity, the connectivity decreases in an unspecific manner, similar to a weight-dependent synaptic scaling, affecting both pre- and postsynaptic elements.

\begin{figure}
\centering
\makebox[\textwidth]{\includegraphics[scale=1]{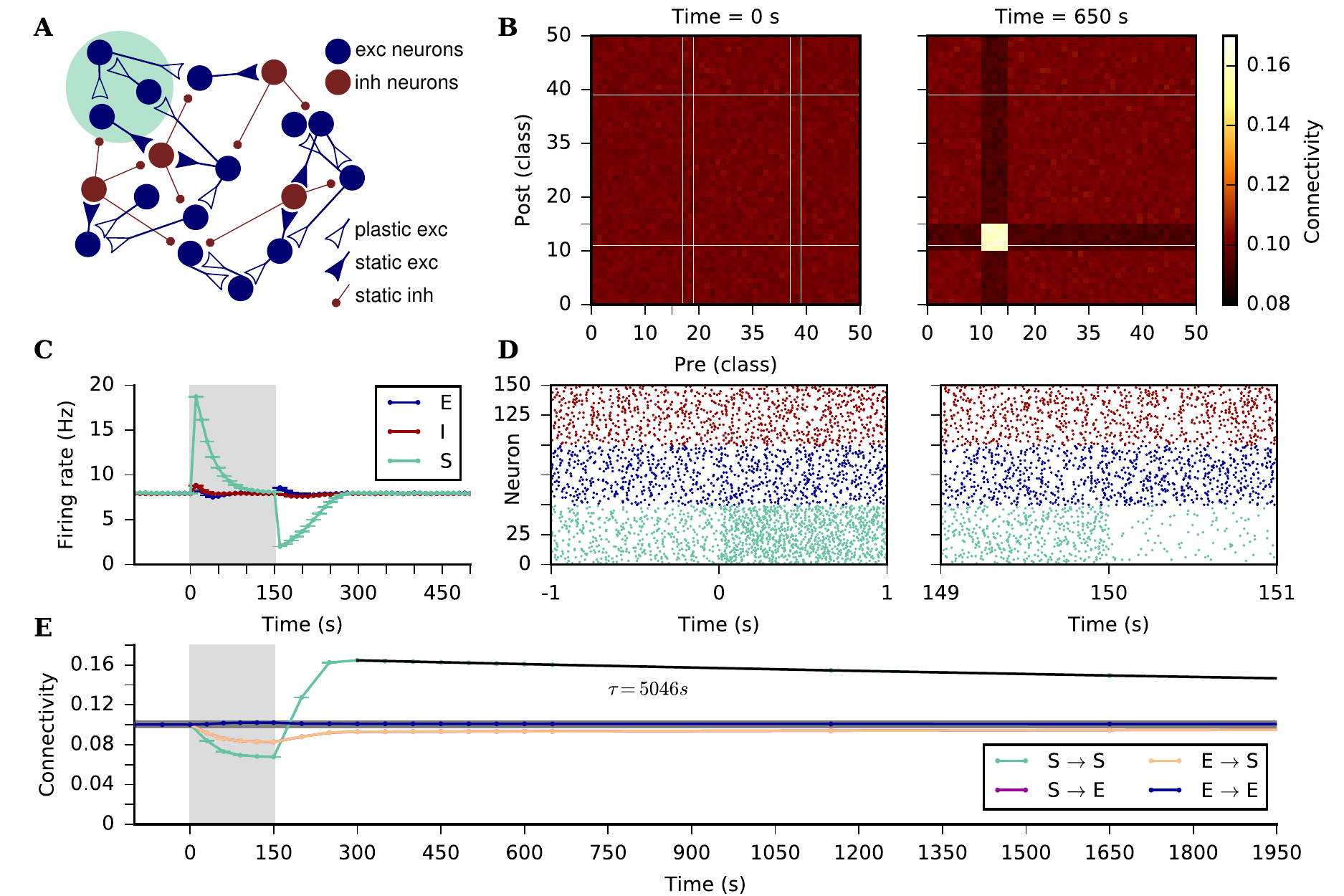}}
\caption{Associative properties of the SP rule. (A)~A recurrent network of excitatory and inhibitory neurons is grown from scratch (see text for details). After it has reached a statistical equilibrium of connectivity, a subgroup (S) comprising $10\%$ of the excitatory neurons is stimulated with a strong external input for $150\,\mathrm{s}$. All the other excitatory (E) and inhibitory (I) neurons in the network are still stimulated with the same external input as during the growth phase. (B)~Connectivity matrix before specific stimulation (left), and after specific stimulation has been off for $500\,\mathrm{s}$ (right). Neurons are divided into $50$ equally large classes and colors correspond to average connectivity between classes. Neurons are sorted such that classes $10$ to $15$ comprise neurons belonging to the stimulated subgroup. (C)~The S neurons (green), E neurons (blue), and I neurons (red) change their firing rates due to a change in external input, but also due to the induced changes in connectivity. As the firing rates of excitatory neurons are subject to individual homeostatic control, they are all back to normal after another $150\,\mathrm{s}$ once the extra stimulus is turned off. Dots and bars indicate mean $\pm$ standard deviation across $10$ independent simulation runs. Highest standard deviation in the time series is $0.09\,\mathrm{Hz}$. (D)~Raster plot for $50$ neurons randomly chosen from S, $50$ from E, and $50$ from I. Shown are~$2\,\mathrm{s}$ before and after specific stimulation starts (left) and $2\,\mathrm{s}$ before and after specific stimulation ends (right). (E)~Average connectivity within the stimulated subgroup (green), among excitatory neurons not belonging to the subgroup (blue), as well as across populations from non-stimulated excitatory neurons to the stimulated subgroup (orange) and from the stimulated subgroup to non-stimulated excitatory neurons (purple). Dots and bars indicate mean $\pm$ standard deviation across $10$ independent simulation runs. Highest standard deviation in the time series is $5 \times 10^{-4}$.  The grey horizontal line indicates the average connectivity right before specific stimulation starts. The black line is an exponential fit to the subgroup to subgroup connectivity, from which the time constant $\tau$ was extracted. The structural association among jointly stimulated neurons induced by stimulation persists for a very long time. Grey boxes in (C) and (E) indicate the time when external input to the stimulated subgroup is on.}
\label{fig:associative_properties} 
\end{figure}

The associative properties of SP are tested by stimulating a subgroup of neurons and quantifying the changes in their connectivity.
After the network was grown and all excitatory neurons were firing roughly at their target rate, a subgroup of the excitatory neurons was stimulated with higher external input for a duration of $150\,\mathrm{s}$.
Upon stimulation onset, the instantaneous firing rate of the stimulated subgroup increases (Fig.~\ref{fig:associative_properties}C), pushing the neurons away from their target rate and triggering a rewiring of their dendrites and axons.
The connectivity of the neurons then decays (Fig.~\ref{fig:associative_properties}E) until the instantaneous rate again reaches its target value, and the average connectivity then stabilizes.
When specific stimulation stops, and all excitatory neurons receive again un-tuned external input, the firing rate of the subgroup drops to a level below the set point, as the excitatory amplification by the recurrent network is now reduced due to the deletion of connections during stimulation.
The firing rate below the set point triggers once again rewiring, and the neurons create new pre- and postsynaptic elements.
Since the effect of specific stimulation on the firing rate of the remaining network is much smaller (Fig.~\ref{fig:associative_properties}C and E), there are more synaptic elements available from the subgroup, and it is more likely for them to create connections within the subgroup than with neurons outside of it (Fig.~\ref{fig:associative_properties}B).

The structural plasticity rule is continuously remodeling the network and there are still changes in connectivity even after the firing rate of the subgroup reaches its target (approximately $300\,\mathrm{s}$ after the end of specific stimulation, see Fig.~\ref{fig:associative_properties}C).
In this case, however, all neurons are firing on average at their target rate, and rewiring is slower, as the change in the number of elements is proportional to $r(t)-\rho$.
The higher connectivity that was formed within the stimulated subgroup lasts for a longer period after activity is back to what it was before stimulation (Fig.~\ref{fig:associative_properties}C and E).

\subsection*{SP leads to feature-specific connectivity in a simple model for the maturation of V1}
\begin{figure}
\centering
\makebox[\textwidth]{\includegraphics[scale=1]{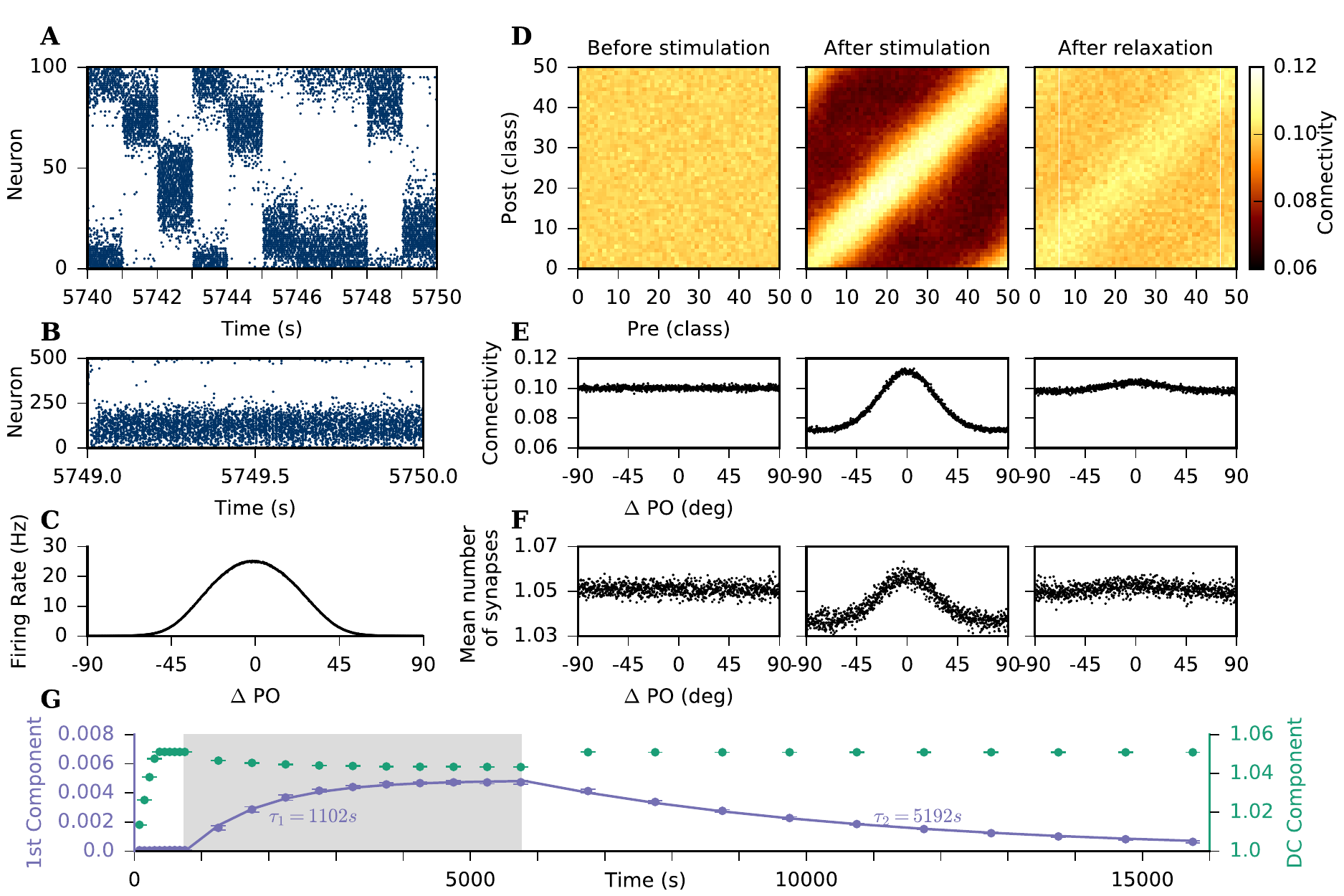}}
\caption{Emergence of feature specific connectivity. (A)~Raster plot of the activity of $100$ randomly chosen excitatory neurons during $10\,\mathrm{s}$ of stimulation. Neurons are sorted according to their input PO. Every $1\,\mathrm{s}$ a new stimulus orientation is randomly chosen and presented, leading to external inputs to all excitatory neurons which was modulated according to their respective input PO. (B)~Activity of $500$ randomly chosen excitatory neurons for the last $1\,\mathrm{s}$ of a stimulation of total duration of $5\,000\,\mathrm{s}$. Neurons are sorted according to their respective input PO. (C)~Tuning curves averaged across all excitatory neurons using the spikes generated during the last $20$ stimuli. (D)~Connectivity matrix, pre- and post-synaptic neurons are sorted according to their PO and subdivided into $50$ equally large classes of similar PO. Colorbar shows average connectivity between classes accounting for multiple contacts. (E)~Mean connection probability, also accounting for multiple contacts, plotted against the difference between pre and post PO. Pairs of neurons are sorted into $1\,000$ bins, and shown is the average connectivity for each individual bin. (F)~Mean number of synapses plotted against the difference between pre and post PO. Only contacts between pairs of neurons that contain at least one synapse are considered. (D-F)~Left column: Connectivity at the end of the initial growth phase ($t=750\,\mathrm{s}$). No orientation bias in connectivity is visible. Middle column: At the end of the stimulation phase ($t=5\,750\,\mathrm{s}$), connectivity is strongly modulated according to the difference between pre and post PO. Right column: After $10\,000\,\mathrm{s}$ of unmodulated input ($t=15\,750\,\mathrm{s}$), connectivity has still a slight orientation bias. Note that the scales are the same across columns. (G)~First two Fourier components of the connectivity as a function of the difference between pre and post PO (E). The left axis (purple) shows the time evolution of the first component, and the right axis (green) shows the time evolution of the DC component. We use non-linear least squares to fit exponential functions to each time series and extract their time constants, $\tau_{1}$ and $\tau_{2}$, respectively. The grey shaded area indicates the period of visual stimulation. Dots and bars indicate mean $\pm$ standard deviation across $10$ independent simulation runs. Highest standard deviation in the time series is $2.6 \times 10^{-4}$ for the DC component and $2 \times 10^{-4}$ for the first component.} 
\label{fig:sensory_maturation} 
\end{figure}

We were then interested in testing the associative properties in a biologically more realistic scenario.
As an example, we consider a simple model for the maturation of V1, in which different stimuli are presented consecutively, and neurons respond to them according to their own functional preferences.
Once the network was formed and is in equilibrium, excitatory neurons were driven by external input that was tuned to stimulus orientation to simulate visual experience.
Each neuron received external input as a Poissonian spike train, the rate of which was modulated according to a tuning curve with an input PO ($\theta_\mathrm{PO}$) that was randomly assigned at the beginning of the simulation. 
More specifically, the modulation depended on the difference between $\theta_\mathrm{PO}$ and the stimulus orientation (SO), which changed randomly at fixed time intervals.
According to the stimulation protocol, neurons receive a slightly higher external input when the presented SO is similar to their input PO, which also entails a higher output rate (Fig.~\ref{fig:sensory_maturation}A-C).

Before oriented stimulation, the connectivity between excitatory neurons was random, and in particular, there was no bias of connectivity with regard to the PO of neurons (Fig.~\ref{fig:sensory_maturation}D and E left columns).
After the presentation of $5\,000$ different stimuli, however, the connectivity pattern of the excitatory neurons change, and neurons are more likely to connect to neurons with similar PO (Fig.~\ref{fig:sensory_maturation}D middles column).
The connectivity is, therefore, modulated according to the difference in the POs of pairs of neurons (Fig.~\ref{fig:sensory_maturation}E middle column).
During modulated stimulation, neurons fire either at a lower or at a higher rate than their target rate, according to the difference between the SO and their PO.
Whenever they fire lower than their target rate, they create synaptic elements. In contrast, if they fire higher than their target rate, they delete synaptic elements.
Neurons with similar PO increase their number of elements at the same stimulus periods and, therefore, have a higher probability of creating synapses between each other.

Fig.~\ref{fig:sensory_maturation}D and E refer to average connectivity between classes of neurons, and synapses contribute the same to connectivity irrespective of whether they are established between two different pairs of neurons or between the same pair.
Fig.~\ref{fig:sensory_maturation}F show how the number of synapses between pairs of connected neurons are modulated according to the difference in their respective POs.
After visual stimulation, neurons are more likely to create multiple synapses to other neurons with similar PO, as compared to neurons with different PO.
Another aspect to notice on Fig.~\ref{fig:sensory_maturation}D and E is the conspicuous drop in average connectivity following the feature-specific stimulation.
One possible explanation for this drop is the non-linearity of the input-output response curve of LIF neurons.
During oriented stimulation the average output is higher, although the average input to the neurons is still the same as before stimulation.
We performed simulations with a static network with $10\%$ connectivity and the same modulated external input, mimicking visual experience. 
We found that the average firing rate of the neurons increased by $1\,\mathrm{Hz}$ during stimulation (see Supplementary material).
In the network with SP such an increase in firing rate would lead to a decrease in average connectivity to keep average firing rate at its target value.

\subsection*{Structural changes induced by stimulation decay very slowly}
As in the case of subgroup stimulation (Fig.~\ref{fig:associative_properties}E), the network structure formed during visual stimulation persists long after the end of stimulation.
After $5\,000$ stimuli, we set the external input to all excitatory neurons back to a uniform value of $\nu_\mathrm{ext}$ to see what happens to the SP induced connectivity in absence of any structured stimulation.
Even after $10\,000\,\mathrm{s}$ of non-modulated external input, the connectivity of excitatory neurons is still slightly modulated according to their PO (Fig.~\ref{fig:sensory_maturation}D and E right columns).

Fig.~\ref{fig:sensory_maturation}G shows the time evolution of the first Fourier components of the feature connectivity modulation (Fig.~\ref{fig:sensory_maturation}E) during network growth, stimulation and post-stimulation phase.
As previously observed in Fig.~\ref{fig:sensory_maturation}D and E, there is a drop in average connectivity during the visual stimulation phase, which can be seen in this case as a decrease in the DC component of the modulated signal.
Also, during the visual stimulation phase, there is an increase in the first Fourier component, corresponding to the modulation of the signal that happens simultaneously with the decrease in average connectivity (Fig.~\ref{fig:sensory_maturation}G).

After the stimulation phase, when all excitatory neurons receive once again the same non-modulated external input $\nu_\mathrm{ext}$, the average connectivity returns quickly to the value it had before stimulation.
The modulation component of the connectivity, on the other hand, decays slowly back to its original value, indicating that there are different time scales for creating and destroying feature specific connectivity based on the modulation of external input.
We used non-linear least squares to fit an exponential function to the first Fourier component of the modulated signal during and after stimulation and extracted the time constant for both process, shown in Fig.~\ref{fig:sensory_maturation}G.
As described previously, the process of creating feature specific connectivity is $5$ times faster ($\tau_{1} \approx 1\,000\,\mathrm{s}$) than destroying it ($\tau_{2} \approx 5\,000\,\mathrm{s}$).
The time series of the Fourier components for multiplicity modulation are qualitatively very similar to the feature specific modulation (see Supplementary material). 

\section*{Discussion}
In our study, we employed a structural plasticity (SP) rule based on the homeostasis of firing rates \cite{Butz2013,Diaz-Pier2016} to grow random networks of excitatory and inhibitory neurons. 
We allowed only EE connections to grow, all other synapses were static.
We showed that in such a configuration an implementation of homeostatic structural plasticity shares important functional properties with a Hebbian learning rule, with different time scales for the creation and decay of the newly formed associations.
In a generic model of visual cortex, we show that feature-specific connectivity, similar to what has been observed in V1 of mice, can emerge from the SP model.
This feature-specific connectivity persists even after the feature-specific stimulation has been turned off.

The time scales of structural plasticity in the brain cover several orders of magnitude, ranging from minutes \cite{Bonhoeffer1999} up to hours or even days \cite{Harris1999a,Hofer2009,Minerbi2009}.
Here we show how the SP model can be adjusted to allow investigations stable through time scales covering at least three orders of magnitude. 
Minerbi \textit{et al.} \cite{Minerbi2009} continuously imaged cultures of rat cortical neurons and found that, although the distribution of synaptic sizes was stable over days, individual synapses were continuously remodeled.
In the SP model, individual pre- and postsynaptic elements are remodeled on a fast time scale, whereas the global network structure evolves at a much slower pace.
Therefore, average connectivity and the distributions of in- and outdegrees might be stable, but individual synapses are still plastic, continuously creating new and deleting existing synapses.

We show in Fig.~\ref{fig:time_scale} that the time scale of network growth depends on the growth parameter for the synaptic elements ($\beta$).
A thorough study on the influence of all parameters of the growth rule on network remodeling was beyond the scope of our present study.
Therefore, we used the simplest combination of identical linear controllers for both pre- and postsynaptic elements.
There is, however, experimental evidence that the growth of axons is somewhat slower than that of spines \cite{Majewska2006}, but the exact rules governing the growth of synaptic elements are still unknown.
The choice of a linear function to implement the homeostatic controller of firing rates is, in any case, in accordance with empirical studies that demonstrated an increase in the number of newly formed spines on cortical neurons in adult mice after monocular deprivation \cite{Hofer2009} and after small lesions of the retina \cite{Keck2008}.
More recently, several studies \cite{Hengen2013,Keck2013,Barnes2015} showed a regulation of firing rate through homeostatic plasticity in the rodent visual cortex \textit{in vivo}.
Keck \textit{et al.} \cite{Keck2013} demonstrated an increase of spine size \textit{in vivo} after a change of sensory input through a retinal lesion, indicating a compensatory recovery through functional plasticity, but no change in spine density after the lesion.
This does not completely discard the hypothesis that sensory deprivation could trigger structural plasticity mechanisms, as the homeostatic regulation of activity is probably the result of an interaction between functional and structural plasticity.
In this specific case, the spine turnover due to structural plasticity as previously observed by Keck \textit{et al.} \cite{Keck2008}, and an increase of the strength of existing connections together lead to a recovery of neuronal activity.
Regarding presynaptic elements, Canty \textit{et al.} \cite{Canty2013} have recently demonstrated axon regrowth \textit{in vivo} after ablation, with axonal bouton densities similar to the state before the lesion, in accordance to a putative homeostatic mechanism.

Several known aspects of cortical network structure and dynamics were not reproduced in our simulations, such as a broad and skewed distribution of firing rates \cite{Hromadka2008,OConnor2010,Mizuseki2013} and synaptic strength \cite{Cossell2015}, and a specific motif statistics for pairs and triplets of neurons \cite{Song2005,Perin2011}.
The structure and dynamics observed in cortical neuronal networks, however, emerge from the interplay of multiple plasticity mechanisms.
A full account of structural remodeling of cortical networks should, therefore, include multiple plasticity mechanisms \cite{Toyoizumi2014,Deger2017}.
With such models, however, is not an easy task to understand what are the effects of individual processes, and what are the effects of combined mechanisms.
On one side, by simulating only one plasticity rule in isolation, we were able to report a very interesting property of a growth rule based on firing rate of homeostasis.
On the other side, it must remain open how SP interacts with other forms of plasticity, and how the above-mentioned property is changed in the presence of other plastic mechanisms that simultaneously update connectivity in the network.

We have provided support for an idea put forward by Dammasch \cite{Dammasch1989}: Hebbian plasticity is not necessarily tied to individual synapses, but can also emerge as a system property.
The homeostatic control of structural plasticity is achieved on the level of whole neurons, and not of individual synapses.
In this model, neurons have control over the number of synaptic elements, which are putative synapses, but not directly over the formation of specific synapses.
The realization of a synaptic contact is, in fact, implemented by randomly wiring available elements.
Thus, Hebbian learning is implemented through the availability and random wiring of free elements in the network.
In contrast to traditional rules implementing Hebbian learning at individual synapses, in the SP model it is not necessary that the neurons keep track of the individual activity of other neurons.
Instead, they only need to keep track of their own activity, and a random wiring scheme implements the correlation dependence.
Hebbian association, therefore, is formed due to neurons controlling their own total input and output, but not the weight of individual synapses, an idea that may be related to the neurocentric view of learning proposed by Titley \textit{et al.} \cite{Titley2017}.
In their review, Fauth \& Tetzlaff \cite{Fauth2016} distinguish two types of structural plasticity rules: (i)~Hebbian, if there is an increase (decrease) of the number of synapses during high (low) activity and (ii)~homeostatic, if there is an increase (decrease) of the number of synapses during low (high) activity.
The SP rule is, according to this definition, a specific variant of homeostatic structural plasticity.
On the network level, however, it implements a form of Hebbian plasticity.
The classification proposed by Fauth \& Tetzlaff \cite{Fauth2016} takes the rules causing the changes in number of pre- and postsynaptic elements into consideration.
Another possible classification, however, would consider the effects of the plastic mechanisms on the connectivity.
It is actually not an easy task to distinguish these two options in experiments, since what we observe is the effect on the network, and it may be impossible to know what were the mechanisms that led to the observed effects.

Another interesting feature implemented by SP in our specific study are the different time scales for establishing and deleting modulated connectivity.
If we consider specific non-random connectivity to implement some sort of memory for previous experiences of the system, this would mean that the system learns faster than it forgets.
This appears to happen because the rewiring in the network is triggered by (and depends on) the discrepancy between the actual activity of the neuron and its target rate.
Learning takes place if there is modulated external input, and the neuronal firing rate is drawn away from its setpoint, leading to strong rewiring.
When the external input is not modulated any more, the neuron's activity recovers quickly back to its target rate, and the rewiring then becomes very slow, depending on the amplitude of random fluctuations.
In a more theoretical framework, Fauth \textit{et al.} \cite{Fauth2015a} showed recently that fast learning and slow forgetting can occur in a stochastic model of structural plasticity.
In their model, new synapses are randomly formed with a constant probability, and randomly deleted with a probability that depends on the number of the existing synapses and the current stimulation.
In our simulations, in contrast, both the creation and the deletion of synapses depend on the number of synaptic elements of each neuron, which in turn depends on its level of activity.
Deletion also depends on the number of existing synaptic contacts between pre- and postsynaptic neurons, due to competition when deleting an existing contact.
Different rules for the growth of synaptic elements could of course lead to different dependencies.
These rules might also influence other properties of the SP model we describe in this paper, such as the capability to form associations.

Hiratani \& Fukai \cite{Hiratani2014} demonstrated the formation of cell assemblies of strongly connected cells in a random recurrent network with short-term depression, log-STDP \cite{Gilson2011} and homeostatic plasticity.
Similarly to other computational models \cite{Sadeh2015c,Zenke2015}, the stronger connectivity is accompanied by sustained activity of neurons, consistent with the concept of a working memory.
Our results, in contrast, show a memory trace in the connectivity of neurons without sustained activity, more consistent with the idea of contextual memories.
Another clear difference between these models concerns the time evolution of average synaptic weights.
Hiratani \& Fukai \cite{Hiratani2014} show an increase in average synaptic weight throughout the stimulation.
In our simulations, the average connectivity decreases during stimulation time due to the homeostatic principles underlying the plasticity rule, and it increases after the specific stimulation has been turned off.
This would imply that immediately after the end of stimulation, connectivity between the stimulated neurons is lower than baseline.
Although this seems counter-intuitive, there have indeed been studies showing perceptual deterioration after trial repetition for subjects tested in a certain task on the same day \cite{Mednick2005,Censor2006,Ofen2007,Mednick2003}, followed by perceptual improvement after $24$ and $48$ hours \cite{Mednick2003}, which would be in agreement to the observed dynamics of connectivity in our simulations.
Experimental data on plasticity usually report values of connectivity before and after a stimulation, but do not allow insight into the connectivity dynamics.
Knowing the time evolution of these connectivity values during different stimulation protocols could give us important hints about the exact mechanisms of plastic changes and help constrain the plasticity models.

A straight-forward consequence of homeostatic plasticity is the stabilization of activity in neuronal networks. 
This aspect has been thoroughly studied over many years [see \cite{Turrigiano2012} for a review on homeostatic plasticity for stabilizing neuronal activity].
The synaptic homeostasis hypothesis should be mentioned in this context \cite{Tononi2014}. 
It states that Hebbian learning during awake states leads to an increase in firing rates, and homeostatic plasticity during sleep states has the goal to restore activity back to baseline levels.
Hengen \textit{et al.} \cite{Hengen2016} recently showed that the opposite is the case.
They continuously monitored the firing rate of individual visual cortical neurons in freely behaving rats over several days and showed that homeostasis is actually inhibited by sleep and promoted by wake states.
It is of course possible that homeostatic plasticity has an exclusive role for network stabilization, even if it is active during wake and not sleep states.
In any case, all these aspects taken together suggest that there could be more to homeostatic plasticity than just stabilizing the network.



\section*{Acknowledgements}
Supported by Erasmus Mundus / EuroSPIN, BMBF (grant BFNT 01GQ0830) and DFG (grant EXC 1086). The HPC facilities are funded by the state of Baden-W\"urttemberg through bwHPC and DFG grant INST 39/963-1 FUGG. We thank Sandra Diaz-Pier and Mika\"el Naveau from the Research Center J\"ulich for support on new features of NEST, and Uwe Grauer from the Bernstein Center Freiburg as well as Bernd Wiebelt and Michael Janczyk from the Freiburg University Computing Center for their assistance with HPC applications. The article processing charge was covered by the open access publication fund of the University of Freiburg.

\section*{Author contributions statement}
J.G. and S.R. conceptualized the main goals, J.G. performed the simulations, J.G. and S.R. analyzed the results, S.R. supervised the work, J.G. and S.R. wrote and revised the manuscript.

\section*{Competing interests}
The authors declare no competing interests.

\end{document}